%% file: Berger_March_2012.tex
\documentclass[authoryear,preprint,review,12pt]{elsarticle}

 \pdfoutput=1
\usepackage{graphicx, amsmath, amssymb, fancyhdr,bm,xspace}
\input{definitions}

\newcommand{\bmath}[1]{\ensuremath{\bm{#1}}\xspace}

\renewcommand\vec[1]{ {\bmath #1} }
\newcommand\stress[2]{{\mathrm \sigma}_{#1 #2}}
\newcommand\stresstensor{{\bmath \sigma}}
\newcommand\straintensor{{\overline{\bmath \epsilon}}}
\newcommand\strain[1]{\epsilon_{#1 }}
\newcommand\load[1]{ \vec{F}_{\mathrm{#1}} }

\newcommand\xav[1]{ \left\langle #1\right\rangle_x }
\newcommand\yav[1]{ \left\langle #1\right\rangle_y }
\newcommand\zav[1]{ \left\langle #1\right\rangle_z }
\newcommand\iav[1]{ \left\langle #1\right\rangle_i }
\newcommand\xextent{ \xav {L_x} }
\newcommand\yextent{ \yav {L_y} }
\newcommand\zextent{ \zav {L_z} }

\newcommand\meanstress[2]{ \overline {\stress #1 #2} }
\newcommand\meanstrain[2]{ \overline {\strain {#1 #2}} }
\newcommand\nin[1]{N_{#1}^\mathrm{in}}
\newcommand\nout[1]{N_{#1}^\mathrm{out}}
\newcommand\flux[1]{\Phi_{\mathrm {#1}}}

\begin{document}

\begin{frontmatter}

\title
{The flow of forces through cellular materials}
%{Detecting auxetic behaviour using stress flow vectors}
%{Deriving the Poisson's Ratio for Cellular Materials from Load Path Analysis}
\author{Mitchell A. Berger}
%\runningauthor{Berger} \runningtitle{The flow of forces through cellular materials}

 %\address{ CEMPS, U. of Exeter, EX4 4QF U.K. \\ \url{m.berger@exeter.ac.uk} \\
  %           }

\address{CEMPS, U. of Exeter, Exeter  EX4 4QF U.K. \\ \textit{m.berger@exeter.ac.uk}\\
\center{25March 2012}}%

\begin{abstract}

Describing and measuring the elastic properties of cellular materials such as honeycombs and foams can be a difficult problem when the cell structure is disordered. This paper suggests that tracking the flow of forces through the material can help in visualizing and understanding how the geometry of the cell structure affects the elastic behaviour. 
The mean strain tensor for a sample of material can be calculated by summing over the force paths, weighted by the strengths of the paths. This method emphasizes the paths with the greatest stress, which can have the most dynamic effect. The path averaging technique reproduces previous expressions for the Poisson's ratio of regular honeycombs, but easily extends to disordered honeycombs and foams.

\end{abstract}
\end{frontmatter}

%\maketitle
\tableofcontents

\section{Introduction}
Strain is a measure of how a material deforms when subject to stresses. For homogeneous isotropic materials the relation between stress and strain can be parameterized by elastic moduli such as the Young's modulus and the Poisson's ratio. When stretched in one direction most materials deform in the transverse directions. Poisson's ratio measures the relative amounts of strain in the parallel and transverse directions and is defined as positive for most materials where stretching in one direction causes contraction in the other two.

Cellular materials such as honeycombs and foams can exhibit \emph{auxetic} \citep{evans91} elastic properties, in particular negative Poisson's ratios \citep{lakes87,evans89,masters96,gibson97,grima06,scarpa08}. For example, detailed modeling of regular 2 dimensional molecular honeycombs \citep{evans94} and 3-dimensional foams \citep{evans95} with re-entrant cell geometries find auxetic behaviour. Auxetic materials have many applications, including impact resistance and sound and vibration damping \citep{alderson07}. 

Disordered materials can also be auxetic, but their tensorial description is complicated by the inherent inhomogeneities and anisotropies \citep{Horrigan09}. An approximate continuous description can be made using mean field theory \citep{gaspar03, koenders05, gaspar08}. Most materials exhibit some defects; thus even for highly ordered materials it is important to understand how inhomogeneities affect the elastic properties.

We model a cellular material as a network of elements connected together at nodes. For two-dimensional honeycombs, the elements might be beams representing the edges of cells. Foams have three dimensional cells; these can be open (as in sponges) or closed. However, because of surface tensions during manufacture or growth, most of the mass density resides in the edges where faces intersect \citep{gibson82}. Thus many three dimensional foams can still be modeled as a network of beams connected at nodes. We assume the network is in equilibrium with no internal forces. We then analyze the elastic properties of the network when subject to motions such as hinging which make little change to the lengths of the beams. 

The principal mathematical objects of elasticity theory, stress and strain tensors, are locally defined. Most experimental procedures, however, involve some sort of non-local averaging. For example, many studies employ the \emph{engineering strain} $\overline{\epsilon_{ij}}$, which for a rectangular material sample gives the ratio of the average displacement in the $i$ direction to the length of the region in the $j$ direction.

For a cellular material the microscopic stresses and strains vary strongly on the scale of a few cells.  Thus an appropriate theoretical description of the physics at the cellular scale (or larger) should also involve averaged quantities. This paper reports on the properties of a variant of the engineering strain, where quantities are averaged over all paths taken by the stresses through the material, weighted by the strengths of the paths. This method also provides simple interpretations of some basic theorems regarding the mean stress tensor.

Section 2 describes the flow of forces through cellular materials. Sections 3 and 4 give expressions for mean sizes and displacements in a material sample, based on averaging over force paths. Section 5 discusses the calculation of the path averaged strain tensor. Example calculations will be presented in section 6. In particular, the method of path averaging gives  Poisson's ratios for regular honeycombs identical to those found in the literature \citep{masters96, gibson97,scarpa08}. Section 8 gives conclusions. 

\begin{figure}%
\begin{centering}
\includegraphics[width=12cm]{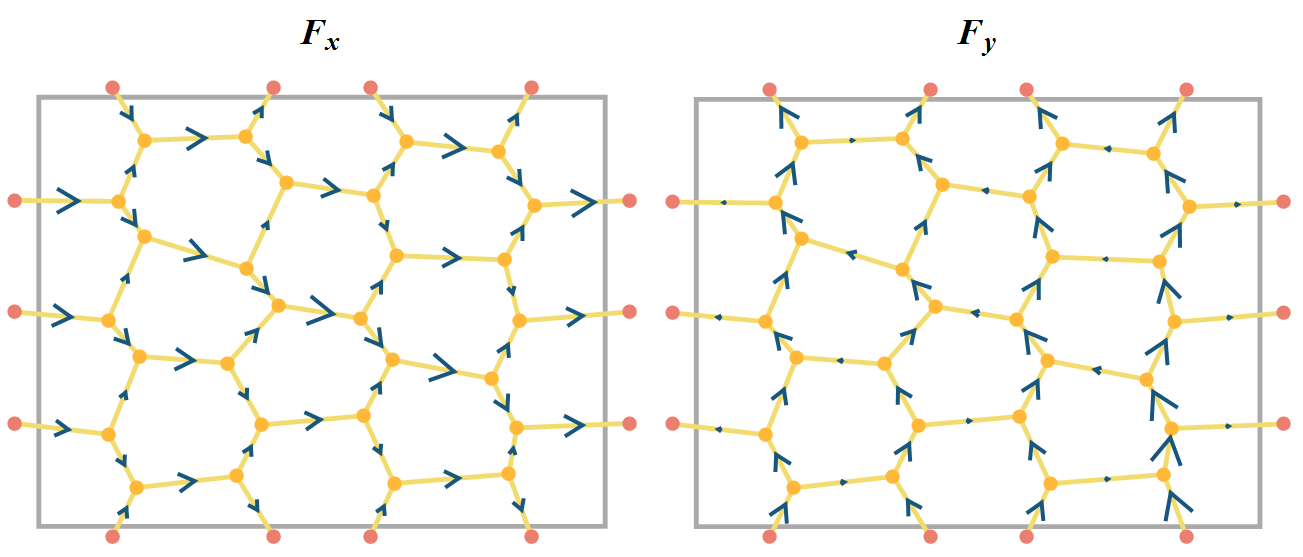}\break%
\includegraphics[width=12cm]{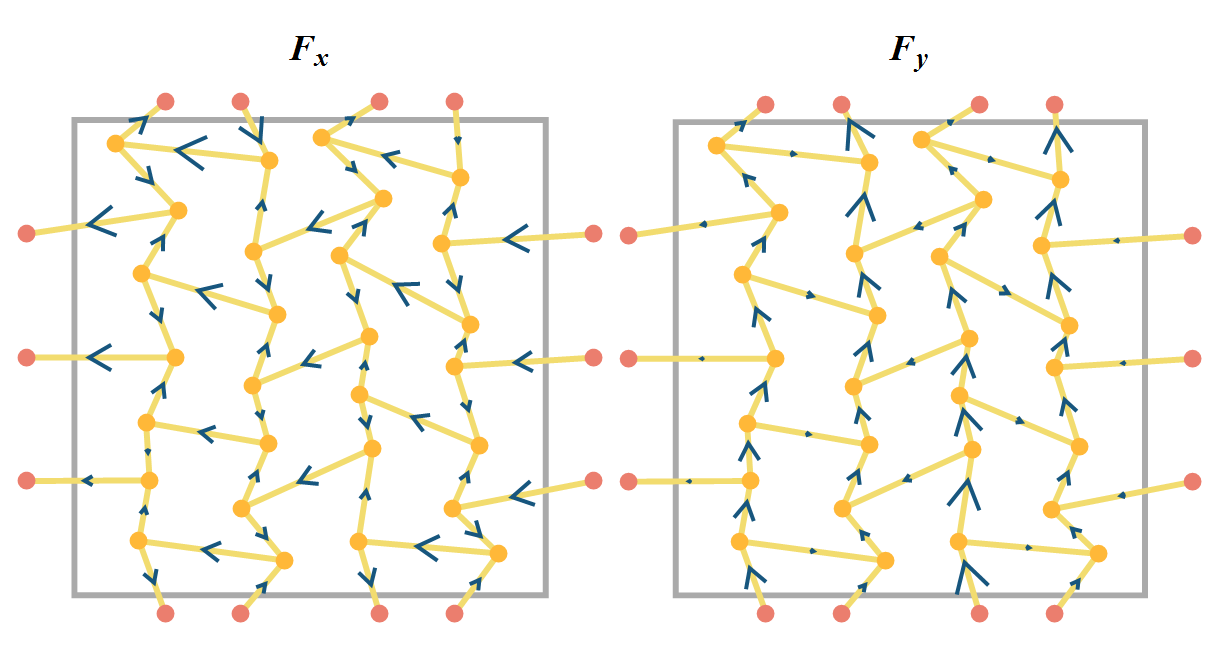}%
\caption{The flow of forces in disordered honeycombs. Section \ref{sec:ex} describes the initial setup and boundary conditions for these equilibrium configurations. The left diagrams show the directions of $\load x$, while the right diagrams show $\load y$. The size of the arrows is proportional to the strength of the forces. The equilibrium configurations are initial stressed by a small stretch in the vertical direction. Top: A honeycomb with positive Poisson's ratio $\nu_{12} = 0.774$, using  \eref{eq:twod3}. The flow of both $\load x$ and $\load y$ forces are carried  by beams under tension, as the directions are rightwards and upwards. 
Bottom: A honeycomb with negative Poisson's ratio $\nu_{12} = - 0.605 $. Here the $\load y$ forces are carried principally by beams under tension, while the $\load x$ forces are carried principally by compressed beams.}%
\label{fig:loadflows}%
\end{centering}
\end{figure}

\section{Force flows through cellular materials}
The stress tensor $\stress ij$ determines how forces propagate through a material. In the absence of body forces the stress has zero divergence (here repeated indices are summed and $\partial_i = \partial/\partial x_i$):
\begin{equation}
\partial_i \stress ij = 0, \qquad j = 1,2,3.
\end{equation}
Letting $\xhat$ be a unit vector in the $x$ direction, we can then construct a divergence free vector field
\begin{equation}
\load x =  \xhat\cdot\stresstensor ; \qquad {\load x}_i = \stress 1i.
\label{eq:defn}
\end{equation}
Similarly,
$\load y = \yhat\cdot\stresstensor $ and $ \load z = \zhat\cdot\stresstensor$.
The integral curves of $\load x$, $\load y$, and $\load z$ can be used to track the paths of $x$, $y$, and $z$ forces through a medium. These curves are consistent with the definition of load path given in \cite{kelley1995}. Kelley and Elsley point out that the load path defined this way differs from simply following the principal stress direction. This difference becomes strong when substantial shear stresses are present.

Note that $\load x$, $\load y$, and $\load z$ can be read off of rows (or columns) of the stress tensor. The rows of a tensor are not vectors -- but $\xhat\cdot  \stresstensor $ is a vector, so the latter form needs to be used in coordinate transformations. 

At first sight, one might expect a $180^\circ$ ambiguity in the direction of a load path vector. For example, consider a beam aligned in the $\xhat$ direction squeezed from both sides (subject to axial compression). There is no asymmetry between left and right; yet the vector $\load x$ will point to the left. Here the choice of dotting the stress tensor with $\xhat$ rather than $-\xhat$ determines the direction of $\load x$.  As $\stress 11$ is negative for compression, the load flow vector $\load x$ points in the negative $x$ direction. Under tension $\load x$ points to the right. Similarly, a vertical beam under tension has $\load z$ pointing upwards, etc.

Consider the forces on a network of interconnecting beams, as in \fref{fig:loadflows}. Tracking the flow of forces through the beams is analogous to tracking the flow of water through a network of pipes. At each node the x-forces (for example) balance; some of the beams connected to a node will have $\load x$ pointing into the node, others out of the node. The net strength of $\load x$ on incoming beams equals the net strength of $\load x$ on outgoing beams, just as in a water network the net flow on inflow pipes equals the net flow on outflow pipes. While we can consider the flow of forces in any coordinate system, we shall later pay special attention to coordinates aligned along the principal axes of the mean stress tensor.

Suppose the network is enclosed within a boundary surface with outward normal $\nhat$. 
A subset of the beams cross the boundary. Consider a beam labelled $a$ which passes through the boundary at position $\vec x(a)$, with orientation along the unit vector $\rhat (a)$. The beam has cross-sectional area $A$. This beam transfers a net $x$-force $\flux x(a)$ through the boundary equal to the flux of the load vector $\load x(a)$:
\begin{eqnarray}
\flux x(a) &=& \left|\int_a   \stress 1j\hat n_j\da\right| =
\left|\int_a   \load x(a)  \cdot \nhat (a)\da\right| \\
 &=& A | \load x(a)  \cdot \rhat (a)|.
\end{eqnarray}

We assume that external forces enter and exit the network only through these beams. By force balance, the sum of the external forces is zero.
Thus the net flux of $\load x$ through the boundary vanishes:
\begin{equation}
\oint \load x \cdot \nhat  \da = 0.
\end{equation}
As we have a discrete system, we can write this equation as a sum over the boundary beams (see \fref{fig:loadflows}). Call the beam $a$ an \emph{inflow} beam for $x$ forces if $\load x(a) \cdot \nhat  < 0$, or an \emph{outflow} beam if $\load x(a) \cdot \nhat > 0$. For example, if external forces stretch the network in the $x$ direction, so that most of the network is in tension, then inflow beams will be on the left, while outflow beams will be on the right.
Suppose there are $\nout x$ outflow beams, and $\nin x$ inflow beams. Then by overall force balance
\begin{equation}
\sum_{a = 1}^{\nin x} \flux {x}(a) = \sum_{b = 1}^{\nout x} \flux {x}(b)\equiv \flux x. 
\end{equation}
Here the net $x$-force $\flux x$ equals the net inflow (as well as the net outflow). 

The forces entering the system at the inflow beams take multiple paths through the network before exiting through the outflow beams. Let $\flux x(a,b)$ be the net amount of $x$-force which passes from beam $a$ to one particular outflow beam $b$. Then we have
\begin{equation}
\flux {x}(a) = \sum_{b = 1}^{\nout x} \flux {x}(a,b);\qquad \flux {x}(b) = \sum_{a = 1}^{\nin x} \flux {x}(a,b). 
\end{equation}

\section{The size of a network averaged over force paths}

Consider a particular load path taken by the $x$-force $\load x$. Various numbers can be calculated characterizing the geometry of this path, such as its net size in a particular direction. We can then average these numbers over all load paths. A well-known theorem (e.g. \cite{landau86}, ch. 1.2) relates the average stress in a material to a boundary integral over external forces. This theorem will be recast in terms of average path sizes and net fluxes of the force vectors.

We will denote an average over $\load x$ paths by
the symbol $\xav {\,}$. In this section we consider the mean size in the $x$ direction; the next section discusses how this mean size changes during a  deformation.

\subsection{The mean path size}

A path from inflow beam $a$ to outflow beam $b$ travels a net distance in the $x$ direction
\begin{equation}
L_x(a,b) = x(b)-x(a).
\label{eq:lx}
\end{equation}
%similarly $L_y = y(b)-y(a)$ and $L_z = z(b)-z(a)$. 
We can define the average $x$ distance travelled by the $x$ force by summing over all paths the force takes through the network, weighting by the flow strength:
\begin{equation}
\xav {L_x} \equiv \frac 1 {\flux x} \sum_{a = 1}^{\nin x}\sum_{b = 1}^{\nout x}\flux {x}(a,b) L_x(a,b).
\label{eq:meansize}
\end{equation}

For a rectangular sample of material aligned with the axes, all paths which stretch from the left side to the right side should have the same net length $L_x$ in the $x$ direction.
So why should we seek a path-averaged length like $\xav {L_x}$?
First, we will need these expressions when relating the mean stress tensor to a suitably defined path-averaged strain. Secondly,  force paths do not always stretch from one side of a rectangular box to the opposite side. For example, some x-force paths starting on the left side (carrying a perpendicular external load) may end on one of the sides $y = $ constant or $z = $constant. The x-force crossing the boundary at one of these sides connects to an external shear or frictional force.  
Third, quantities like $\xav {L_x}$ provide a measure of average size for irregularly shaped samples. 

The definition of mean size can be rearranged into a boundary sum: 
\begin{eqnarray}
{\flux x}\xav {L_x} & = &  \sum_{a = 1}^{\nin x} \sum_{b = 1}^{\nout x}\flux {x}(a,b)(x(b)-x(a))\\
& = &    \sum_{b = 1}^{\nout x} x(b)  \sum_{a = 1}^{\nin x}\flux {x}(a,b) - \sum_{a = 1}^{\nin x}x(a) \sum_{b = 1}^{\nout x}\flux {x}(a,b)\\
& = &  \sum_{b = 1}^{\nout x} x(b)\flux {x}(b) - \sum_{a = 1}^{\nin x}x(a)\flux {x}(a).
\label{eq:lxx}
\end{eqnarray}
We next relate this expression to the mean stress.

\subsection{Boundary integrals}

The mean stress satisfies a useful relation: The stress averaged over an area (or volume in 3D) is equal to a boundary integral over external forces \citep{landau86}. Let $\vol$ be the volume (or area in 2 dimensions) of the sample of material. Then the mean stress tensor is defined as
\begin{equation}
\meanstress ij \equiv \frac 1 {\vol} \int \stress ij \dv.
\label{eq:2}
\end{equation}
As $\stress ij$ is divergence free, 
\begin{eqnarray}
\meanstress ij  & = & \frac 1 {\vol}\int\stress ik\pder{x_j}{x_k} \diff\vol = \frac 1 {\vol}\oint (\stress ik x_j)n_k\da\\
& = &\frac 1 {\vol}\oint ({\vec F}_i \cdot \nhat) x_j\da .\label{landau}
\end{eqnarray}

%Given that the $b$ beams are outflow beams (with $\load x$ pointing out of the boundary) and the $a$ beams are inflow beams,
%we then have

Let us look at the $11$ component of  \eref{landau}. 
At an inflow rib $a$ (integrating over the rib cross-section) 
\begin{equation}
\int (\load x \cdot \nhat) x\da = -\flux x (a) x(a)
\end{equation}
while at an outflow rib $b$
\begin{equation}
\int (\load x \cdot \nhat) x\da = \flux x(b) x(b).
\end{equation}
Thus
\begin{equation}
 \meanstress 11 = \frac 1 {\vol}\sum_{b = 1}^{\nout x} x(b)\flux {x}(b) - \frac 1 {\vol}\sum_{a = 1}^{\nin x}x(a)\flux {x}(a).\label{landau3}
\end{equation}
We can write this in terms of the path-averaged $\xextent$ as in \eref{eq:lxx}:
\begin{equation}
\meanstress 11 = \frac {\flux x} {\vol} \xextent. \label{meanstx}
\end{equation}
Similarly
\begin{equation}
\meanstress 22 = \frac{\flux y}{\vol}\yextent; \qquad \meanstress 33 = \frac{\flux z}{\vol}\zextent.\label{meanstyz}
\end{equation}

\section{Path-averaged displacements}

We can treat extensions of a network during a deformation in the same way. Suppose that position $\vec x$ changes to $\vec x + \vec u$. Then the size $L_x(a,b)$ of a path will 
change by an amount $\delta L_x(a,b) = u_1(b)-u_1(a)$. We can average this extension over all paths as before:
\begin{equation}
\xav{\delta L_x}\equiv \frac 1 {\flux x} \sum_{a = 1}^{\nin x}\sum_{b = 1}^{\nout x}\flux {x}(a,b)\delta L_x(a,b).
\label{eq:meandeltaxa}
\end{equation}
We can rearrange this as in \eref{eq:lxx}, then relate it to the boundary stress:
\begin{eqnarray}
{\flux x}\xav{\delta L_x}& = &\sum_{b = 1}^{\nout x} u_1(b)\flux {x}(b) - \sum_{a = 1}^{\nin x}u_1(a)\flux {x}(a)\\
& = &\oint (\load x \cdot \hat n) u_x \da.
\label{eq:meandeltax}
\end{eqnarray}

In general we can define the tensor
\begin{equation}
T_{ij} \equiv \half\int \stress ik \partial_k u_j\dv.
\label{eq:itensor}
\end{equation}
As the stress is divergence-free, this becomes
\begin{eqnarray}
T_{ij} & = &\half \oint n_k \stress ik  u_j\da = \half\oint (\load i \cdot \hat n) u_j \da\\
& = & \half{\flux i}\iav{\delta L_j}.
\label{eq:itensor2}
\end{eqnarray}

\section{The path averaged strain and Poisson's ratio}
We will use our path-averaged extensions and displacements to define a mean strain. First, we define the mean strain tensor
$\straintensor$ in a coordinate system aligned with the principal axes of the mean stress tensor. If desired, $\straintensor$ can be transformed into any other coordinate system. The principal axis system provides the most natural context for defining force paths and the fluxes $\flux x$, $\flux y$, and $\flux z$ employed in the weighted averages. Quantities in the principal axis system will be denoted by a $*$.

Let $x$, $y$, and $z$ be principal axis coordinates. Then we define
\begin{equation}
{\strain 1}^* \equiv \meanstrain 11 \equiv\frac{\xav {\delta L_x}}{\xav {L_x}},
\label{eq:diagstrain}
\end{equation}
and similarly for ${\strain 2}^*$ and ${\strain 3}^*$. 

Note from \eref{meanstx} that (letting    
$\sigma_1^* \equiv  {\meanstress 11}$, etc.)
\begin{equation}
 {\strain 1}^* =  \frac{\flux x \xav {\delta L_x}}{\vol{\sigma_1}^*}\qquad  
 {\strain 2}^* =  \frac{\flux y \yav {\delta L_y}}{\vol{\sigma_2}^*}\qquad
 {\strain 3}^* =  \frac{\flux z \zav {\delta L_z}}{\vol{\sigma_3}^*} .
  \label{eq:strain1}
\end{equation}

The general strain tensor, valid in any coordinate system, with these quantities as diagonal elements in the principal axis system is
\begin{equation}
\meanstrain ij\equiv  \frac 1 {2\vol}\left(({\meanstress ik})^{-1}
T_{kj} + ({\meanstress jk})^{-1}
T_{ki} \right).
\label{eq:straindef}
\end{equation}

The Poisson's ratio in the $x$ and $y$ directions constructed from these quantities is
\begin{equation}
\nu_{12}^* \equiv - \frac{{\strain 2}^*}{{\strain 1}^*}.
\label{eq:poisson}
\end{equation}

\section{Examples}
\subsection{Three beams connected to one node}\label{triplets}

\begin{figure}%
\begin{centering}
\includegraphics[width=5cm]{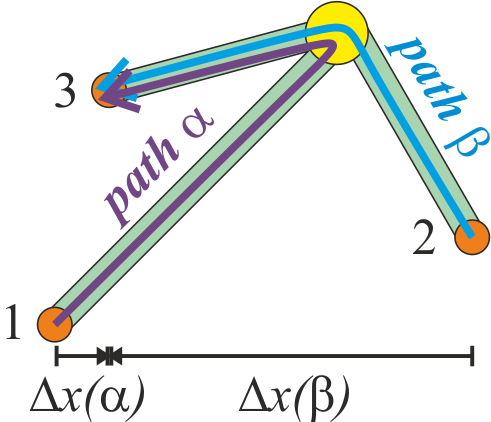}\hskip 1cm\includegraphics[width=7cm]{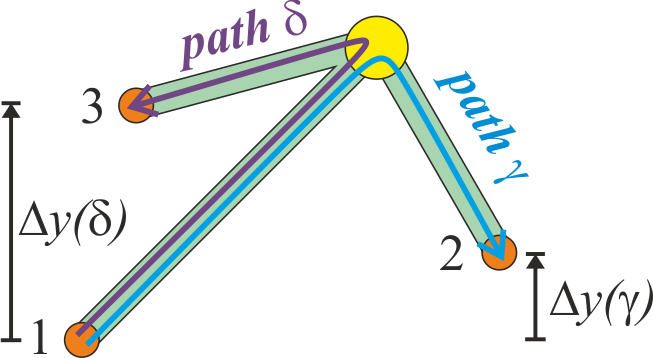}%
\caption{Left: The flow of the $x-$force vector $\load x$ through three connected beams. There are inflows along beams 1 and 2, and an outflow along beam 3. Right: The flow of the $y-$force vector $\load y$ through the node. There is an inflow along beam 1, and outflows along beams 2 and 3. For this particular example, the lengths are proportional to 
$(\ell_1, \ell_2, \ell_3) =  (27,16,15)$, and the orientations are $(\phi_1, \phi_2, \phi_3) = (-3\pi/4, -\pi/3, -7\pi/8)$. The Poisson's ratio defined by \eref{eq:poisson} is $\nu_{12} =- 0.178$.
The path-averaged horizontal extent is
${\flux x} \xextent  \equiv  \flux {x\,(\text{path $\alpha$})}\Delta x_{(\text{path $\alpha$})}+\flux {x\,(\text{path $\beta$})}\Delta x_{(\text{path $\beta$})} $.
Similarly, in the vertical direction 
${\flux y}\yextent  \equiv  {\flux {y\,(\text{path $\gamma$})}\Delta y _{(\text{path $\gamma$})}+\flux {y\,(\text{path $\delta$})}\Delta y _{(\text{path $\delta$})}} 
$. 
}%
\label{fig:jelly}%
\end{centering}
\end{figure}

The simplest non-trivial network consists of three beams connected to one node (see figure \ref{fig:jelly}). By force balance, the three beams must lie in a plane.
Such triply connected nodes form the basic building blocks of many two dimensional honeycombs (e.g. hexagonal honeycombs). 

Let the three beams have lengths $2\ell_1$, 2$\ell_2$, and 2$\ell_3$ and cross-sections $A_1$, $A_2$, and $A_3$. The half-lengths $\ell_1$,$\ell_2$, and $\ell_3$ give the extent of the beams in the neighbourhood of the node. The beams have orientations $\phi_1$, $\phi_2$, and $\phi_3$ with respect to the $x$ axis. The strength of beam 1 is $\tau_1$, etc. Force balance requires
\begin{eqnarray}
\tau_1 A_1 \cos \phi_1 + \tau_2 A_2\cos\phi_2 + \tau_3A_3\cos \phi_3 & = & 0;\\
\tau_1 A_1\sin \phi_1 + \tau_2 A_2\sin\phi_2 + \tau_3A_3\sin \phi_3 & = & 0.
\label{eq:forcebalance}
\end{eqnarray}
%These equations are equivalent to
%\begin{eqnarray}
%\frac{\tau_1}{\sin(\phi_2-\phi_3)} = \frac{\tau_2}{\sin(\phi_3-\phi_1)}=\frac{\tau_3}{\sin(\phi_1-\phi_2)}.
%\label{eq:taus}
%\end{eqnarray}
%Thus if one $\tau$ is specified to give the overall strength at the node, then the other two are also specified.
%Let
%\begin{equation}
%c_1 \equiv \cos\phi_1;\qquad \sin\phi_1 \equiv \sin\phi_1; \qquad \text{etc.}
%\label{eq:cone}
%\end{equation}

The mean stress tensor is
\begin{equation}
\vol\overline \stresstensor = \ell_1 A_1\stresstensor_1 + \ell_2 A_2 \stresstensor_2 + \ell_3A_3 \stresstensor_3; \qquad \stresstensor_i ={\tau_i}  \begin{pmatrix}
	&\cos\phi_i^2 &\cos\phi_i\sin\phi_i\\ &\cos\phi_i \sin\phi_i &\sin\phi_i^2
\end{pmatrix}.
\label{eq:means}
\end{equation}

\subsection{The regular hexagonal honeycomb}

Consider a regular hexagonal honeycomb with beam 1 horizontal ($\phi_1 = 0$) and beams 2 and 3 at equal and opposite angles $\phi_2=-\phi_3 \equiv \phi$ with respect to the $x$ axis.
%(see figure \ref{fig:honey}).
Let $\ell_1=h/2$ and $\ell_2 = \ell_3 = \ell/2$. 
To maintain force balance at the node, $\tau_2A_2 = \tau_3A_3 = \tau_1A_1/(2 \cos\phi)$. 
We average the stress over the neighbourhood of the node, which includes half of the lengths of each beam. Thus
\begin{equation}
\vol \overline \stresstensor  =\frac{\tau_1A_1}{2}\begin{pmatrix}
	h + \ell\cos\phi &0\\
	0 &\ell\sin^2\phi/\cos\phi
\end{pmatrix}. 
\label{eq:bees}
\end{equation}
The ratio of the diagonal terms is
\begin{equation}
\nu_{12}=\frac {\meanstress 11}{\meanstress 22} = \frac{\cos\phi((h/\ell) + \cos\phi)}{\sin^2\phi}=\frac{\sin\theta((h/\ell) + \sin\theta)}{\cos^2\theta},
\label{eq:stress ratio}
\end{equation}
identical to previous expressions for Poisson's ratio  in regular honeycombs \citep{masters96,gibson97,scarpa08}
%\cite{masters96} 
for flexed or hinged ribs.

\subsection{Disordered honeycombs}\label{sec:ex} 

A simple numerical model illustrates the use of path averaging for disordered two dimensional networks. We start with a regular honeycomb residing inside a rectangular region, and perturb the position of each node with a random step taken from a Gaussian distribution. The elements between nodes are assumed to carry a linear (Hooke's law) force. The outermost nodes are then moved outwards or inwards slightly, and the equations of motion are solved with heavy damping until the system settles into an equilibrium state. The mean stress tensor can now be calculated, giving a prediction of the Poisson's ratio of any small motions to or away from this equilibrium. To obtain the displacement, we move the top and bottom boundary nodes a tiny amount, while letting the nodes on the left and right boundaries move freely.

%Figure \ref{fig:jelly} shows the flow of forces through  networks derived from perturbing  honeycombs of regular hexagons.
%The discrepancy between left and right sides of  \eref{eq:twod3} was of order the size of the strain, and is due to the finite size of the motion. For the top example, the cells started with a regular hexagon, $\nu_{12} = 1$, and were perturbed with random steps (with a root-mean-square stepsize of $0.1 h$).

To illustrate for a reentrant honeycomb, we began with the cells in a regular hexagon pattern with $\ell=h/2$, $\phi = 0.65\pi$, and  $\nu_{21} = -0.884$, and perturbed with random steps (rms $0.1 h$) (see bottom of \fref{fig:loadflows}). A stretch in the vertical direction of $0.1 h$ was followed by a relaxation to an equilibrium. The equilibrium configuration had $\nu_{21}=- 0.605$. The inclination of the principal axis frame was negligible: only 0.02 degrees. The individual nodes can be analyzed individually, as in section \ref{triplets}. Here the average $\nu$ of the individual nodes was $-0.566$ in their principal axis systems, not much different from the total configuration. However, the principal axes of individual nodes were often misaligned with respect to the total: in the $xy$ frame, Poisson's ratios varied from -4.28 to -0.45. The Poisson's ratio calculated without weighting by path strength was $0.595$, only two percent off from the path-averaged ratio.

For the top example, the cells started as regular hexagons stretched in the horizontal direction, with $h=1.5 \ell$, $\phi = \pi/3$, and  $\nu_{21} = 0.75$, before the random perturbation. The Poisson's ratio became $\nu_{21}=0.774$ after a random perturbation. The inclination of the principal axis frame was 15 degrees. The average $\nu_{21}$ of the individual nodes was $0.685$. The Poisson's ratio calculated without weighting by path strength was $0.903$.

% Using ordinary averaging (without weighting by the forces) to obtain the engineering strain gave $\nu = ?$.

\section{Conclusions}
The elastic properties of a disordered cellular material depend on the size and geometry of the sample being measured.
To describe these elastic properties, we need to employ some method of averaging in order to 
obtain effective elastic moduli. This paper has examines the possibility of averaging over all paths taken by the forces through the material. This method emphasizes the paths which carry the most stress, and hence have the most influence on the mechanics of the material. Also, in many cases it may give effective moduli close to those calculated by cruder averaging.

This result has immediate implications for engineers wishing to design an auxetic material. Consider a honeycomb stretched in the $y$ direction, so that the mean $\meanstress 22 <0$. Then for $\nu_{12}$ to be negative, we must have a positive $\meanstress 11$,
as in figure \ref{fig:loadflows}. Thus the solution to the force balance equations must provide a mixture of elements where some are compressed and some are under tension. Here the $y$ forces must be primarily carried by the elements under tension, and the $x$ forces carried by the compressed elements. 

%The equilibrium solutions are strongly affected by the initial geometry. 
A reentrant honeycomb geometry \citep{masters96} accomplishes this naturally: while the $y$ force $\load y$ must ultimately flow from bottom to top, the paths zig-zag both upwards and downwards on the way. The downwards sections of the paths are under compression; these contribute to the negative flow of $x$ forces $\load x$. Not all auxetic networks have an obvious reentrant geometry \citep{Horrigan09}; an analysis based on load paths and the mean stress tensor may help to understand auxetic behavior of disordered networks in general.

Furthermore, there may be a hierarchy of structures contributing to auxetic behavior \citep{taylor11}. The methods proposed here
can be measured for any subvolume of a material, rectangular or irregular in shape. This flexibility allows analysis of elastic properties on many length scales.

\section{Acknowledgments}
I am very pleased to thank Ken Evans for introducing me to this subject, and for valuable discussions and encouragement. I also am pleased to thank  Chris Smith and Phillipe Young for many discussions.

\bibliographystyle{model2-names}

\bibliography{AuxeticsBibTeX}

\end{document}

%% file: definitions.tex
\renewcommand\div{{\bf \nabla} \cdot }% RENEW
\newcommand{\xhat}{\hat{\vec x}}
\newcommand{\yhat}{\hat{\vec y}}
\newcommand\zhat{\hat {\vec z}}
\newcommand\nhat{\hat {\vec n}}
\newcommand\rhat{\hat {\vec r}}

\newcommand\half{ \frac{1}{2} }

\newcommand\rmd{ {\mathrm d} }
\newcommand\pder[2]{ \frac{\partial #1}{\partial #2} }

\newcommand\diff[1]{\,\rmd #1}

\newcommand\vol{{\cal V}}

\newcommand\dv{~{\mathrm d}^3 x}
\newcommand\da{~{\mathrm d}^2 x}

\newcounter{problemnumber}

\newcommand\eref[1]{equation~(\ref{#1})}

\newcommand\fref[1]{figure \ref{#1}}